\documentclass[twocolumn,
notitlepage,showpacs,
prb,eqsecnum,superscriptaddress]{revtex4-1}
\usepackage{amsmath}
\usepackage{epsf}
\usepackage{graphicx}
\usepackage{dcolumn}
\usepackage{bm}

\newcommand{\req}[1]{Eq.~(\ref{#1})}
\newcommand{\reqs}[1]{Eqs.~(\ref{#1})}
\newcommand{\rref}[1]{(\ref{#1})}

\renewcommand{\r}{\mathbf{r}}

\newcommand{\beq}{\begin{equation}}
\newcommand{\eeq}{\end{equation}}
\newcommand{\be}{\begin{equation}}
\newcommand{\ee}{\end{equation}}
\newcommand{\beqa}{\begin{eqnarray}}
\newcommand{\eeqa}{\end{eqnarray}}
\newcommand{\bea}{\begin{align}}
\newcommand{\eea}{\end{align}}
\usepackage{amssymb}
\usepackage{array}
\usepackage{amsmath}
\usepackage{graphicx}\usepackage{graphics}
\usepackage{dcolumn}
\usepackage{bm}\usepackage{varioref}

\begin{document}
\def\boldsymbol#1{\mbox{\boldmath$#1$}}

\title{Memory effects, two color percolation, and the temperature dependence of Mott's variable range hopping}

\author{Oded Agam}
\affiliation{Physics Department, Columbia University, New York, NY 10027, USA}
\affiliation{The Racah Institute of Physics, The Hebrew University of Jerusalem, 91904, Israel}

\author{Igor L. Aleiner}
\affiliation{Physics Department, Columbia University, New York, NY 10027, USA}

\date{\today}
\begin{abstract}
  There are three basic processes that determine hopping
  transport: (a) hopping between normally empty sites (i.e. having exponentially
  small occupation numbers at equilibrium); (b) hopping between normally occupied sites, and (c) transitions between normally occupied and unoccupied sites. In
  conventional theories all these processes are considered Markovian
  and the correlations of occupation numbers of different sites are believed to be small
  (i.e. not exponential in temperature).  We show that, contrary to this belief, memory effects suppress the processes of type (c), and manifest
  themselves in a subleading {\em exponential} temperature dependence
  of the variable range hopping conductivity. This temperature dependence
  originates from the property that sites of type (a) and (b) form two
  independent resistor networks that are weakly coupled to each other by
  processes of type (c). This leads to a two-color percolation problem which we
  solve in the critical region.
\end{abstract}
\pacs{72.20.Ee,72.20.-i}

\maketitle

\nopagebreak

\section{Introduction}

The dominant mechanism of transport in disordered systems, when the
carrier's states are localized and the temperature is sufficiently
low, is hopping (for a review see
Ref.~[\onlinecite{ShklovskiiEfros}]). The hopping rate between two
sites, say $i$ and $j$, is governed mainly by two factors: the
tunneling probability, $\exp(-2 r_{ij}/\xi)$, and the Boltzmann
factor, $\exp(-\epsilon_{ij}/T)$, where $r_{ij}$ and
$\epsilon_{ij}$ are the hopping distance and the energy difference
between sites, respectively, while $\xi$ is the localization length,
and $T$ is the temperature. The leading exponential dependence of the
dc conductivity is given by the variable range hopping formula which
may be obtained by maximizing the product of these two factors. Thus
\be
\sigma_{VRH}(T) =B(T) \exp \left[-\left(\frac{T_0}{T}
  \right)^{p}\right]. \label{Main-Old}
\ee
Here $T_0$ is the
characteristic temperature of the hopping mechanism which depends on
the localization length and the behavior of the density of states near
the Fermi level. In particular for the case where long range Coulomb
interactions can be neglected, $p=1/(1+d)$ where $d$ is the effective
dimension of the system \cite{Mott68}, while $p=1/2$, independently of
$d$, if Coulomb interactions are dominant\cite{Efros75}. The
prefactor, $B(T)$, is believed to be a power-law of $T$. The most
elaborate theoretical estimates of $T_0$ are based on mapping the
problem to the random resistor network of Miller and
Abrahams\cite{MA}, and analyzing it using percolation theory (for a
review see Ref.~[\onlinecite{Percolation}]).

However, formula (\ref{Main-Old}) has been obtained under several
simplifying assumptions (see e.g.~discussion in
Ref.~[\onlinecite{Ambegaokar71}]). Among them is the assumption that
memory effects, associated with temporal changes in the occupation of
the sites, can be neglected. Such correlations, in the context
of hopping conductivity, were studied long ago
\cite{Richards77,Levin82,Kurobe82,Thouless89, Gartner92}, but their
implication for the temperature dependence of variable-range
hopping conductivity was overlooked. Similar correlations emerging from on-site Hubbard repulsion have been shown to have negligible effect\cite{Levin82,ShklovskiiEfros}.

Here we show that memory effects
manifest themselves in the appearance of a subleading contribution to
the exponential dependence of the conductivity (\ref{Main-Old}). In
particular, for systems in which long-range interactions can be
neglected a more accurate form of the hopping conductivity is
\be
\sigma(T) =B(T) \exp \left[-\left(\frac{T_0}{T}
  \right)^{1/(d+1)} + \alpha_d \left(\frac{T_0}{T}
  \right)^{\mu_d} \right], \label{Main-new}
\ee
where $\alpha_d$ is a
constant of order unity, while $\mu_2= 25/129 \simeq 0.2$ and
$\mu_3\simeq 0.1$. The second term in the exponent can be
misinterpreted as an anomalously large preexponential factor when
Eq.~(\ref{Main-Old}) is used in order to fit the experimental data.

In the next section we provide a qualitative explanation of this
result. In Sec.~\ref{sec3}, we
formulate the problem, write down the rate equations which govern
the dynamics on the random resistor network,
decouple the lowest order correlation function,
and obtain the effective conductances of the random
resistor network modified by the classical memory effect. Next, in Sec.~\ref{sec4}, we reduce the problem of finding the hopping conductivity
to the problem of "two color percolation" \cite{Ioselevich95,Matveev95}, and show how the solution
of this problem in the critical region (considered improperly in Refs.~\onlinecite{Ioselevich95,Matveev95}) leads to \req{Main-new}. Finally, in Sec.~\ref{sec5}, we conclude and discuss the implications of \req{Main-new}. In this paper, for simplicity, we consider spinless electrons.

\section{Qualitative explanation}
In order to present the qualitative explanation of formula
(\ref{Main-new}) it is instructive, first, to recall the percolation
theory approach to hopping conductivity in the case where memory
effects are neglected \cite{Ambegaokar71, Shklovskii71,ShklovskiiEfros}. In this approach, one considerers the system as a
network of random resistors with the conductances, $G_{ij}$, through which the current flows such that
\be
\begin{split}
& G_{ij}\propto \exp(-\Upsilon_{ij}),
\\
& \Upsilon_{ij}\equiv \frac{2|\r_i-\r_j|}{\xi}+ \frac{ |\epsilon_i|+|\epsilon_j|+ |\epsilon_i - \epsilon_j|}{2T}.
\end{split}
\label{exponent-qualitative}
\ee
Here $\epsilon_i$, and $\r_i$ are, respectively, the energy and the location of the center of mass
of the localized wave-function of site $i$.

To identify the effective conductivity of this network we choose some
threshold conductance $G_*$ and cut out all resistors with lower
conductances $G_{ij}<G_*$. At the same time we shortcut all resistors of higher
conductance. In this process we obtain an effective resistor network
that with an exponential accuracy has the same resistance as the original
network. Now let us start with a low value of $G_*$ and begin
raising it up such that an increasing number of resistors are cut
out. Passing some value of $G_*$ the infinite cluster of
connected resistors will break up into separate finite-size
clusters. The value where this transition takes place,
$G_*=G_c$, is the critical point of percolation transition
and the exponential temperature dependence of the conductivity of the network (\ref{Main-Old}) is the same as the
exponential temperature dependence of $G_c\simeq \exp(-\Upsilon_c)$, with  $\Upsilon_c= (T_0/T)^{1/(d+1)}$.

\begin{figure}
\begin{center}
\includegraphics[width=\columnwidth]{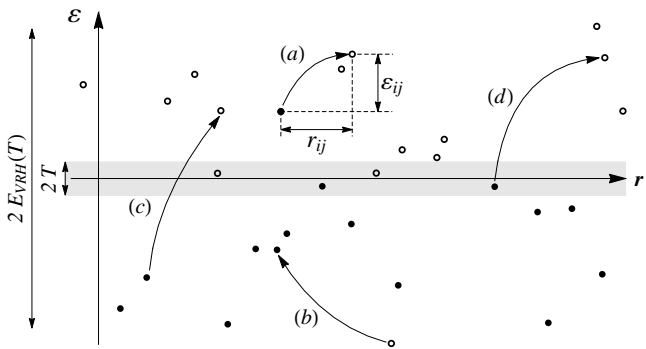}
\caption{Illustration of the processes that take place in hopping conductivity. The typical transitions are over distances much larger than the localization length, $r_{ij}\gg \xi$, and over energy differences larger than the temperature $E_{VRH}(T) \gg T$. These include:(a) hopping between two sites with equilibrium occupation numbers that are exponentially small; (b) hopping between two sites whose equilibrium occupation numbers are exponentially close to 1, and (c) hopping between two states located on either sides of the chemical potential. The latter transitions are strongly suppressed due to memory effects. The memory effect for transitions of the type denoted by (d) are limited, but these transitions are rare due to small phase-space volume.}
\label{fig1}
\end{center}
\end{figure}

The observable conductivity is determined by the conductors forming the bottlenecks of the percolation cluster such that $\Upsilon_{ij}\simeq \Upsilon_c \gg 1$.
An important consequence of the percolation picture is that at low
enough temperature the transport is dominated by hops of typical distances, $r_{ij} \sim L_{VRH}=\xi \Upsilon_c$,
 much larger than the localization length $\xi$, and typical
energies $|\epsilon_i| \sim E_{VRH}(T) =T \Upsilon_c$ much larger than the temperature, as illustrated in Fig.~\ref{fig1}.
Because the site energies are much higher than the temperature,$|\epsilon_i|\gg T$, we can distinguish
three main types of hops that are involved in variable-range
hopping: (a) An electron above the chemical potential hops to an empty
state which is also above the chemical potential $\epsilon_{i},\epsilon_j>0$ (electron-like process);
(b) A hole below the chemical potential hopes to another site below the chemical potential $\epsilon_{i},\epsilon_j<0$  (hole-like processes);
(c) An electron below the chemical potential $\epsilon_i<0$ hops to an empty state
above the chemical potential $\epsilon_j>0$ (creation of an electron-hole pair). Other processes, e.g. an electron within the
energy band $T$ near the chemical potential that jumps to some other empty
states [such as (d) in Fig.~\ref{fig1}], have small probability due to small phase space volume.

If one ignores  memory effects (the Markovian approximation), all three processes described above
lead to the same conductance $G_c$. Therefore sites above
and below the chemical potential form a single percolating
cluster. The Markovian approximation amounts to the assumption that the
equilibrium values of the occupation numbers are restored immediately
after the particle hops from its original site to another
site, and the same process repeats.

\begin{figure}
\begin{center}
\includegraphics[width=\columnwidth]{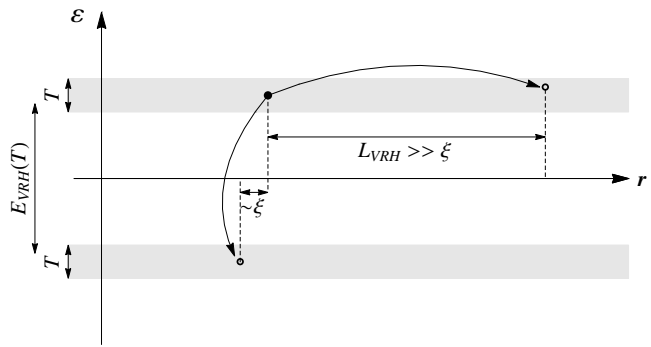}
\caption{An illustration of the importance of memory effect for process of type (c) shown in Fig.~1. After an electron-hole pair is created, the electron can either jump to some unoccupied site or recombine with the hole left behind. The latter process is much more probable, see explanation below Eq.~(\ref{distances}).}
\label{fig2}
\end{center}
\end{figure}

However, in strongly disordered insulators, relaxation to equilibration
is rather slow because it is determined by the very same hopping processes. Therefore, one cannot neglect the correlations that
are generated between occupation numbers of different sites due to
the dynamics of the charge carriers. The role of those correlations for processes of type (c) is very different from that
for the processes of type (a) and (b). To illustrate this difference in the most simple way, let us consider the electron and hole sites in the energy intervals $|\epsilon_i\pm E_{VRH}/2|\lesssim T$, as illustrated in Fig.~\ref{fig2}. It can be seen that the typical hopping distance for the critical resistors (those that belong to the percolating network) is:
\be
\begin{split}
&|\r_i-\r_j| \simeq \Upsilon_c\xi/4\gg \xi ;  \ \epsilon_i,\epsilon_j > 0 \text{   or } \epsilon_i,\epsilon_j < 0;
\\
&|\r_i-\r_j| \simeq \xi;  \ \epsilon_i > 0;\ \epsilon_j < 0,
\end{split}
\label{distances}
\ee
whereas the conductances are essentially the same. Now, let us assume that a hopping process occurred and focus our attention on the next hopping of the same electron (or hole) to other sites.
If the process is electron like (hole like) the next hop occurs with equal probabilities to all connected empty (occupied) sites without
 reference to the memory regarding the initial state [in this case we have essentially a one-particle random walk because $\exp(-|\epsilon_i|/T)$ means the probability to excite such a particle]. However, the situation is different for electron-hole creation with the thermal exponent $\exp(-|\epsilon_i-\epsilon_j|/T)$. Let us consider the fate of the electron created after such a process. It has the possibility to diffuse away via electron-like processes on the percolation cluster
or return and recombine with the hole. The hopping rates to diffuse away are exponentially small, $\propto \exp(-2|\r_i-\r_j|/\xi)\propto\exp(-\Upsilon_c/2)$, whereas the probability to recombine does not contain such exponential smallness [see \req{distances}]. As a result the resistors of type (c), being nominally the threshold resistors, do not contribute to the transport.

From these considerations it follows that to the leading approximation in $T/T_0$ the transport in hopping systems takes place on two parallel and independent percolation networks, one associated with electrons above the chemical potential and the other associated with holes below the chemical potential. This picture of independent clusters gives back \req{Main-Old} albeit with a different value  $T_0 \to 2T_0$. This change,
by itself, is not interesting because the parameter $T_0$ is usually determined from a fit of the measured conductivity to formula (\ref{Main-Old}), and there is no independent experiment allowing the direct  extraction of $T_0$.
\begin{figure}
\begin{center}
\includegraphics[width=0.8\columnwidth]{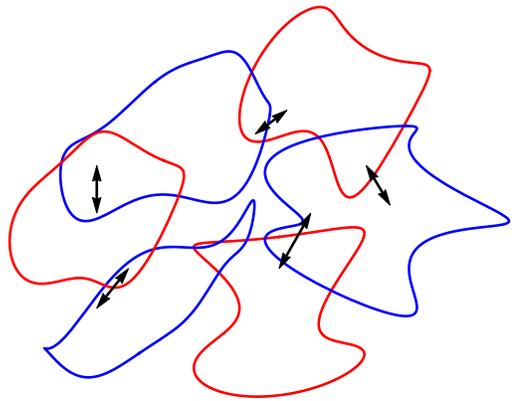}
\caption{(Color online) Illustration of "two color percolation
  problem". Each color represents off-critical clusters of resistor
  networks associated either with states above the chemical potential
  (transitions of type (a) in Fig.~1) or the states below the chemical
  potential (transitions labeled by (b) in Fig.~1). The black arrows
  represent sparse transition between the two groups of clusters due
  to processes where one site is located in a narrow energy band of
  width $T$ near the chemical potential, such as that labeled by (d)
  in Fig.~1. }
\label{fig3}
\end{center}
\end{figure}

The interesting consequence of the memory effect is the subleading exponential dependence in \req{Main-new}.
To explain its origin, let us take into account the resistors which connect the two
percolating networks. These are the resistors for which the relative
change in one of the occupation numbers (that are associated with the
transition) is of order one.  This takes place when the corresponding
site is located within an energy band of order $T$ near the chemical
potential, as illustrated, e.g., in case (d) of Fig.~\ref{fig1}. The
probability of finding such a resistor is small because the phase
space volume of these processes is small. Nevertheless these resistors,
although sparse, restore the single network of resistors, albeit with
a higher conductivity.

To clarify the reason for this increase, let us denote by
$G_c$ the conductance of a typical resistor, at the percolation
threshold, of a single percolation network, either that of the
electrons or that of the holes (we assume particle-hole
symmetry). Now let us increase the conductance $G_c \to
G_c + \delta G$. According to the procedure described
above, this increase results in a breakup of the percolating networks
into finite size clusters, as illustrated in Fig.~\ref{fig3}. But this does not
imply that the total network of resistors becomes disconnected, because there may be enough resistors connecting clusters of the two
groups to form a single infinite cluster. The typical size of the
clusters reduces with the increase of $\delta G$, and the new
conductance threshold can be identified by increasing $\delta G$
up to the point where the average number of connecting resistors for
two overlapping clusters (one from each group) is of order one.  From
these considerations it follows that the increase in the conductivity of the network
is temperature-dependent because the probability of finding a
connecting resistor is proportional to $T/E_{VRH}(T)$.  In Sec.~IV we
show how these considerations and the critical behavior of the
the clusters away from the percolation threshold lead to
formula (\ref{Main-new}).

\section{Memory effects and their modification of the resistors network.}
\label{sec3}

\subsection{Dynamics of average occupations and the Markovian resistors network.}
Consider the hopping conductivity of spinless electrons in a disordered system, and let $n_i$ be a random variable
which is one if the $i$-th site is occupied by an electron and zero if it is empty. The rate equation which governs the average of the occupation number at site $i$ is:
\be
\begin{split}
&\frac{d\langle n_i\rangle }{dt}= \sum_{k\neq i}  I_{k\to i};
\\
& I_{k\to i}=
\frac{\langle
  n_k(1-n_i)\rangle}{\tau_{k \to i}}- \frac{\langle
  n_i(1-n_k)\rangle }{\tau_{i \to k}},
\end{split}
\label{Qeq}
\ee
where $\langle \dots \rangle$ denotes the averaging over any probability distribution, and $1/\tau_{i \to j}$ is the hopping rate from site $i$ to site $j$. The first term in the expression for link-current, $I_{k\to i}$, describes the transition of an
electron from an occupied site $k$ to some unoccupied site $i$, while the second term is associated with the reverse process.

In equilibrium, the correlators on the right-hand side of the above equation decouple, and detailed balance ensures that the time derivatives vanish.
In what follows we denote by the equilibrium values of the average occupation numbers
by the superscript $eq$.  Thus detailed balance implies:
\be
\frac{n_k^{eq}(1-n_i^{eq})}{\tau_{k \to i}}= \frac{ n_i^{eq}(1-n_k^{eq}) }{\tau_{i \to k}}.
\label{detailed-balance}
\ee

In order to characterize the behavior of the system when it is out of equilibrium it is convenient to parametrize the occupation numbers by using a new random variable, $\psi_i$, such that:
\be
n_i  =  n_i^{eq} + n_i^{eq}(1-n_i^{eq})\psi_i
\label{parametrization}
\ee
With this  definition Eq.~(\ref{Qeq}) reduces to:
\be
\begin{split}
& n_i^{eq}(1-n_i^{eq}) \frac{ d\langle \psi_i\rangle} {dt} =
\sum_{k} I_{k\to i},
\\
& I_{k\to i}=\frac{\langle \psi_k-\psi_i\rangle+\langle{\cal F}_{ik}\rangle }{\tau_{ik}},\label{qnetwork}
\end{split}
\ee
where
\be
\frac{1}{\tau_{ik}}= \frac{1}{\tau_{i\to k}}n_i^{eq}(1-n_k^{eq})
\label{rates}
\ee
is the relaxation rate between the sites $i$ and $k$, while
\be
{\cal F}_{ik} =(n_k^{eq}-n_i^{eq}) \psi_i \psi_k \label{Fik}
\ee
are random variables which represents ``forces'' that contain the memory information. In the traditional approach of linear response and mean-field theory (where memory effect is neglected) ${\cal F}_{ik}=0$.  This is because ${\cal F}_{ik}$ is quadratic in the $\psi_i=- eV_i/{T}$ variables which describe the out of equilibrium change in the occupation numbers, where $V_i$ is the site potential.  With this approximation
$eI_{i\to k}$  from \req{qnetwork} is, essentially, the same as for the Kirchhoff's law for a network of random resistors with conductances
\be
G_{ik}=\frac{e^2}{T\tau_{ik}}.
\label{conductances}
\ee

The central point of the memory effect is that the non-equilibrium correlations ${\cal F}_{ij}$
arise already within the linear response theory, i.e. $\langle \psi_i \psi_k\rangle \sim \langle \psi_i - \psi_k\rangle$. In what follows, these correlations will be studied in detail.

\subsection{Dynamics of the correlation functions.}

Memory effects generate correlations among occupation numbers at different sites. In order to study these correlations we need their equations of motion. To derive these equations, let us define $P_{i,j,k...}^{(n_i,n_j,n_k...)}$ as the joint probability of having $n_x$ particles at site $x$ where $x=,i,j,k...$ and $n_x$ is a random variable that can be either one or zero. Then clearly $\langle n_i n_j \rangle = P_{i,j}^{(1,1)}$, and therefore in order to characterize the evolution of $\langle n_i n_j \rangle $ it is sufficient to describe the evolution of $P_{i,j}^{(1,1)}$ in time. Due to the Pauli principle this evolution can be associated only with transitions to (or from) other sites i.e.,
\be
\frac{d P_{i,j}^{(1,1)}}{d t} \!=\! \sum_{k\neq, i,j}\left[ \frac{P_{i,j,k}^{(0,1,1)}}{\tau_{k\to i}} \!+\! \frac{P_{i,j,k}^{(1,0,1)}}{\tau_{k\to j}} \!-\!\frac{P_{i,j,k}^{(1,1,0)}}{\tau_{i\to k}}\!-\! \frac{P_{i,j,k}^{(1,1,0)}}{\tau_{j\to  k}}  \right].
\ee
Expressing the probabilities in the above equation in terms of the averages of the occupation numbers [e.g. $P_{i,j,k}^{(0,1,1)}=\langle (1-n_i)n_jn_k \rangle$], one obtains
\be
\frac{d \langle n_i n_j \rangle}{dt}\!=\! \sum_{ k\neq i,j} \!\left[ \frac{ \langle
    n_k(1-n_i)n_j\rangle}{\tau_{k\to i}} \!-\! \frac{ \langle
    n_i(1-n_k)n_j\rangle}{\tau_{i\to k}}\right] \!+ (i \!\leftrightarrow \! j)
\ee

Finally, by using the parametrization (\ref{parametrization}), the above equation reduces to
 \beqa
&& n_i^{eq}(1-n_i^{eq}) n_j^{eq} (1-n_j^{eq}) \frac{ d\langle\langle
  \psi_i\psi_j\rangle\rangle} {dt}  \label{chargecorrelator} \\
  &&~~= \frac{\langle \psi_j-\psi_i \rangle + \langle{\cal
     F}_{ij}\rangle}{\tau_{ij}}\langle {n}_i-{n}_j\rangle \nonumber \\
     && ~~+ \sum_{k\neq i,j} \left[ \frac{(1-n_j^{eq}) n_j^{eq}}{\tau_{ik}} \langle\langle \psi_j ( \psi_k -\psi_i + {\cal F}_{ik} ) \rangle\rangle +(i\leftrightarrow j) \right], \nonumber
\eeqa
where double angular brackets denote the connected part of the correlation function,
 e.g., $\langle\langle \psi_i\psi_j\rangle\rangle =\langle \psi_i\psi_j\rangle-\langle \psi_i\rangle\langle \psi_j\rangle$.

The above equation expresses two-point correlators in terms of three-point correlators, $\langle \langle \psi_j {\cal F}_{ik} \rangle \rangle \propto \langle \langle \psi_j \psi_i\psi_k \rangle \rangle$. Similarly, the equation for three-point correlators involves four-point correlations, and so on. In order to close this hierarchical system of the kinetic equations we use the following assumptions:
\begin{itemize}
\item
(i) $\langle \psi_i\rangle \ll  1$ so that all the terms
of the type $\langle \psi_i\rangle\langle\psi_j\rangle $ can be
neglected;

\item
(ii)  As follows from (\ref{chargecorrelator}),
$\langle\langle\psi_i\psi_j\rangle\rangle \sim \langle
\psi_i-\psi_j\rangle$, therefore these correlators will be kept;

\item
(iii) Third order cumulants will be neglected,
$\langle\langle\psi_i\psi_j\psi_k\rangle\rangle=0$. This is justified when the system is close enough to equilibrium such that the density of excited carriers is very small,
and the main contribution to the memory effect is short ranged.
\end{itemize}
Using definition (\ref{Fik}) and the above approximations we have
\begin{subequations}
\be
\langle{\cal F}_{ij}\rangle \simeq (n_j^{eq}-n_i^{eq})\langle\langle \psi_i \psi_j\rangle\rangle
\ee
\be
\langle\langle{\cal F}_{ij}\psi_k\rangle\rangle \simeq 0
\ee
\label{decoupling}
\end{subequations}
and \req{chargecorrelator} acquires the closed form
\beqa
&& n_i^{eq}(1-n_i^{eq}) n_j^{eq} (1-n_j^{eq}) \frac{ d\langle\langle
  \psi_i\psi_j\rangle\rangle} {dt} \label{chargecorrelators-decoupled} \\
&& ~~~~=
 \frac{\langle \psi_j-\psi_i \rangle -\left(n_i^{eq}-n_j^{eq}\right) \langle\langle
  \psi_i\psi_j\rangle\rangle }{\tau_{ij}}\left(n_i^{eq}-n_j^{eq}\right)
 \nonumber \\
 && ~~~~+ \sum_{k\neq i,j}\left[
 \frac{\langle\langle\left( \psi_k-\psi_i\right)\psi_j \rangle\rangle}{\tau_{ik}}n^{eq}_j(1-n^{eq}_j)
 + (i \leftrightarrow j)
 \right]  \nonumber
\eeqa

This equation can be also interpreted as the result of adding
Langevin forces into the right hand side of Eq.~(\ref{qnetwork}), so that the first
term in the right hand side of Eq.~(\ref{chargecorrelators-decoupled}) is the
correlator of those forces. The third line of the above equation describes
the diffusive transport of the correlator.

Equations (\ref{chargecorrelators-decoupled}) and \rref{qnetwork} for the averages of $\psi$s constitute a closed set of equations for the transport on the Miller-Abrahams network.

\subsection{The effective resistor network}

Although \req{chargecorrelators-decoupled} is linear its solution in a
general form is not instructive. Nevertheless, it can be simplified by noticing that \req{chargecorrelators-decoupled} describes a diffusive motion of the two endpoints of the correlator  $\langle \langle \psi_i \psi_j \rangle \rangle$ (the last term in the equation) with a drain proportional
to $(n_i^{eq}-n_j^{eq})^2/\tau_{ij}$. This drains restricts the dynamics to the close vicinity of the $(ij)$ bond, or in other words the steady state solution of the correlator is determined only by small number of hops to nearest neighbors sites. Thus, \req{chargecorrelators-decoupled} may be solved in perturbation theory in the number of hops. This is the {\em short-range memory approximation} which is extensively discussed in Appendix A.

In order to obtain the leading result for the steady-state solution
of $\langle \langle \psi_i \psi_j \rangle \rangle$ by using the short range memory approximation,
 one can neglect the correlations associated with in all other sites, i.e., $\langle\langle\psi_{k'\neq i}\psi_{k\neq j}\rangle\rangle=0$. This approximation reduces \reqs{chargecorrelators-decoupled} (for the stationary case) to
\beqa
 && \frac{\langle \psi_j-\psi_i \rangle\left(n_i^{eq}-n_j^{eq}\right)}{\tau_{ij}}
=\langle\langle
  \psi_i\psi_j\rangle\rangle\left[
\frac{\left(n_i^{eq}-n_j^{eq}\right)^2}{\tau_{ij}} \right. \nonumber \\
&& ~~~~+ \left.
\sum_{k\neq i,j}\left(
 \frac{n^{eq}_j(1-n^{eq}_j)}{\tau_{ik}}
 + \frac{n^{eq}_i(1-n^{eq}_i)}{\tau_{jk}}
\right)
\right]\label{charge-short-range}
\eeqa
Solving Eq.~(\ref{charge-short-range}) and substituting the result in
Eqs.~(\ref{qnetwork}--\ref{Fik}), yields
\be
e I_{i \to j}=T \langle\psi_j-\psi_i\rangle G_{ij}^{eff}.
\ee
It is equivalent to the
resistance network model where the conductances \rref{conductances}
between the sites are replaced by their effective values (i.e. the conductances which take into account the memory effect):
\be
G_{ij}^{eff}=\frac{G_{ij}\left[g_{i;j}n^{eq}_j(1-n^{eq}_j)+g_{j;i}n^{eq}_i(1-n^{eq}_i)\right]}{\left(n_i^{eq}\!-\!n_j^{eq}\right)^2
\!+\! g_{i;j}n^{eq}_j(1\!-\! n^{eq}_j)\!+\!g_{j;i}n^{eq}_i(1\!-\!n^{eq}_i)
},
\label{ee}
\ee
where we introduce the shorthand notation
\be
g_{i;j}\equiv \frac{1}{G_{ij}}\sum_{k\neq i,j} G_{ik}.
\label{gij}
\ee
Equation (\ref{ee}) reduces to the result of Richards\cite{Richards77} in the case of one dimensional chain with alternating occupation numbers at equilibrium.

We turn now to discuss implication of \req{ee} for the problem of variable-range hopping.
As explained in the qualitative discussion, variable-range hopping corresponds to hopping through
 sites with energies $|\epsilon_i|\gg T$. Therefore the equilibrium occupation probabilities of these sites are approximately $n_i^{eq}\simeq\Theta(-\epsilon_i)+ \mbox{sign}(\epsilon_i) {e^{-|\epsilon_i|/T}}$, where $\Theta(x)$ is the Heaviside step function. For this case \req{ee} gives markedly different results for the case where both sites are above or below the chemical potential $\epsilon_i\epsilon_j>0$, and when the two sites are at different sides of the chemical potential,  $\epsilon_j\epsilon_i <0$. Let us consider these cases separately:
\begin{itemize}
\item (i) For $\epsilon_i\epsilon_j>0$, and $|\epsilon_{i}|,|\epsilon_{j}|\gg T$, Eq.~(\ref{ee}) reduces to
\beqa
&& G_{ij}^{eff}=G_{ij}\left[ 1  \label{ee-exp1} \right. \\ && ~~\left.
- \frac{\left(e^{-|\epsilon_i|/T}-e^{-|\epsilon_j|/T}\right)^2}{\left(e^{-|\epsilon_i|/T}\!-\!e^{-|\epsilon_j|/T}\right)^2\!
+\!g_{i;j}e^{-|\epsilon_j|/T}\!+\!g_{j;i}e^{-|\epsilon_i|/T}
}
\right], \nonumber
\eeqa
and taking into account that the pair $(i,j)$ belongs to the conducting network, so that
$g_{i;j}\simeq 1$, we conclude that the corrections to the conductances due to the memory effect are
exponentially small.

\item (ii) For $\epsilon_i\epsilon_j<0$, and $|\epsilon_{i}|,|\epsilon_{j}|\gg T$, Eq.~(\ref{ee}) reduces to
\be
\begin{split}
G_{ij}^{eff}=G_{ij}
\frac{g_{i;j}e^{-|\epsilon_j|/T}+g_{j;i}e^{-|\epsilon_i|/T}}{1
+g_{i;j}e^{-|\epsilon_j|/T}+g_{j;i}e^{-|\epsilon_i|/T}
}.
\label{ee-exp3}
\end{split}
\ee
For pairs $(i,j)$ belonging to the conducting network
$g_{i;j}\simeq 1$, and  the resulting effective conductance is exponentially small compared to its bare value, $G_{ij}$.
Thus for this kind of resistors the memory effect is crucial in accordance with our qualitative discussion.
\end{itemize}

There are several ways to avoid the exponentially small effective conductances described in case $(ii)$. One possibility is  associated with situations in which $g_{i;j}\simeq e^{|\epsilon_j|/T}$ are exponentially large. From the structure of these quantities [see Eq. (\ref{gij}] it is evident that there are two ways of achieving this goal. One is when $G_{ik}$ is large, namely hopping to the site $k$ is spatially and energetically close to the site $i$. However, this condition is satisfied for a small phase volume $\propto
\left(T/E_{VRH}(T)\right)^{d+1}$, where $d$ is the dimensionality of the system, and $E_{VRH}(T)=T(T_0/T)^{1/(d+1)}$ is the typical energy band available for variable range hopping, see Fig.~1. Alternatively, one can require that $G_{i,j}$  is exponentially small by having the pair $(i,j)$ far from each other, i.e. $G_{ij}=G_{typical}e^{-|\epsilon_j|/T}$. Here there is no constraint on phase volume but it produces an effective conductance which is exponentially smaller than the typical one (without the memory effect), which means that such resistor between sites $i$ and $j$ does not belong to the percolation network.

The most probable situation of having a typical resistor connecting sites from both sides of the chemical potential is when one of the energies, either  $\epsilon_i$ or
$\epsilon_j$, lie within the energy strip of the order of $T$, as illustrated, for instance, by case (d) of Fig.~1. This condition constrains the phase volume to approximately $T/E_{VRH}(T)$  but it is the least costly way.

\section{Two color percolation and the temperature dependence of the Variable Range Hopping.}
\label{sec4}

In this section we use the "two color percolation" approach in order to identify the temperature dependence of the conductivity in the variable range hopping regime when taking into account the memory effects described in the previous section. To begin with let us briefly review the case of a single color percolation which is realized when the memory effect is ignored, and hopping takes place between any two sites, whether above or below the chemical potential.

The bare conductances $G_{ij}$ of the resistor between sites $i$ and $j$, defined by Eqs.~(\ref{conductances}) and (\ref{rates}), take the form $G_{ij}=B_{ij}(T)\exp(-\Upsilon_{ij})$, where $B_{ij}(T)$ is a prefactor which does not contain any
exponential dependence of the temperature, and $\Upsilon_{ij}$ is defined in Eq.~(\ref{exponent-qualitative}).
 Consider now the network of resistors obtained from those resistors for which
\be
\Upsilon_{ij} \leq \Upsilon_*, \label{percolationCondition}
\ee
 where $\Upsilon_*$ is some arbitrary value. If we denote by $f({\bf r}_i,\epsilon_i)$ the distribution function that site $i$ belongs to this network, then this distribution function satisfies the homogeneous integral equation,
\be
 \lambda f({\bf r}_i,\epsilon_i) = \rho\int d\epsilon_j d^d r_j \Theta(\Upsilon_*-\Upsilon_{ij})f({\bf r}_j,\epsilon_j) \label{eigenEq}
 \ee
 where $\rho$ is the density of states, and $\Theta(x)$ is the Heaviside step function. The eigenvalue $\lambda$, on the left hand side of the above equation, has the meaning of the average number of sites that are connected to a given site on the network. Now let us reduce the value of $\Upsilon_*$  up to the point where the resistors network forms a peculating cluster, and denote by $\Upsilon_c(T)$ and $\lambda_d$ the critical value of $\Upsilon_*$ and the corresponding eigenvalue $\lambda$, respectively, thus
\be
\lambda\left(\Upsilon_c(T)\right)=\lambda_d;
\label{invariant}
\ee
The eigenvalue $\lambda_d$ is called the invariant of the $d$-dimensional bond
percolation problem. It is of order one, and to a good approximation independent of the microscopic details of the percolation problem.  The above equation gives a reasonable estimate for the
percolation threshold as it determines the effective connectivity at
small distances, so that the scaling properties of the percolation
cluster are not important (the status of this equation is similar to that in the mean-field treatment of second order phase transitions).
The observable conductivity (with exponential accuracy) is thus
\be
\sigma \propto \exp\left[-\Upsilon_c(T)\right].
\label{observablesigma1}
\ee
This is the traditional formula for the variable range hopping.

However, as explained in the previous section, the memory effect hinders the transitions from states below and above the chemical potential.  Thus to the leading approximation these two groups form two independent and parallel resistor networks. Taking into account the resistors which connect the two groups, the generalization of Eq.~(\ref{eigenEq}) would be\cite{Matveev95}:
\begin{subequations} \label{eee}
\be
 \lambda \left(\begin{array}{c} f_+({\bf r}_i,\epsilon_i) \\ f_-({\bf r}_{i},\epsilon_{i}) \end{array} \right) = \frac{\rho}{2} \int d\epsilon_j d^d r_j {\cal M}_{ij}\left(\begin{array}{c} f_+({\bf r}_j,\epsilon_j) \\ f_-({\bf r}_{j},\epsilon_{j}) \end{array} \right),
 \ee
where
\be
{\cal M}_{ij}= \left( \begin{array}{cc}
  \Theta(\Upsilon_*-\Upsilon_{ij}) & \Delta(\Upsilon_*,\Upsilon_{ij}) \\ \Delta(\Upsilon_*,\Upsilon_{ij}) & \Theta(\Upsilon_*-\Upsilon_{ij}), \end{array} \right),
\ee
\end{subequations}
$f_\pm({\bf r}_i,\epsilon_i)$ are the probability distributions above ($+$) and below ($-$) the chemical potential, while the matrix elements $\Delta(\Upsilon_*,\Upsilon_{ij})$ are associated with transitions between the two groups.  To the leading approximation, i.e. if $\Delta(\Upsilon_*,\Upsilon_{ij})=0$, the problem reduces to two independent percolation problems.

 The two-color percolation problem, which is realized when $\Delta(\Upsilon_*,\Upsilon_{ij}) \neq 0$, has been studied before in the framework of the mean field approximation\cite{Ioselevich95,Matveev95}. In this treatment the correction to $\Upsilon_c(T)$ due to the transitions between the two groups is calculated by requiring that the invariant $\lambda_d$ is not influenced by the perturbation, $\Delta(\Upsilon_*,\Upsilon_{ij})$. However, this approach is based on the assumption
that the system remains in the mean-field region and
therefore it does not take into account the scaling properties of the percolation cluster.
In what follows we show how these properties can be taken into account when finding the correction to $\Upsilon_c(T)$.

Let us focus our attention on the two independent resistor networks, and start adding bonds between the two groups.
As those bonds are rare, their effects becomes important at length scales which are much larger than the length
of a typical variable range hopping distance, $L_{VRH}$. To estimate the effect of these bonds
let us move slightly from the percolation point of a single group by changing
the invariant $\lambda_d \to \lambda_d - \delta\lambda$
where $\delta \lambda \ll 1$.  First order perturbation theory shows immediately that this change amounts to a change $\Upsilon_c \to \Upsilon_c -\delta \Upsilon$ where
\be
\delta \lambda =\delta \Upsilon \frac{\partial \lambda_d}{\partial \Upsilon_c}
\ee
Now, the invariant $\lambda_d$ is approximately the number of states within a $d+1$ dimensional volume:
\be
\lambda_d \simeq \frac{\rho}{2} \int d^d r_{ij} d\varepsilon_{ij} \Theta (\Upsilon_c-\Upsilon_{ij}).
\ee
Therefore
\be
\frac{\delta \lambda}{\lambda_d} =(d+1)\frac{\delta \Upsilon}{\Upsilon_c}
\ee
and accordingly the conductance \req{observablesigma1} becomes
\be
\sigma \propto \exp\left[- \Upsilon_c(T)+\frac{\delta\lambda}{(d+1)\lambda_d} \Upsilon_c(T)\right]
\label{observablesigma2}
\ee

The clusters consisting solely of sites of one group
are now disconnected, however, the size of the typical cluster is still large
\be
L_{corr} =\frac{L_{VRH}}{|\delta \lambda|^{\nu_d}}
\label{corrlength}
\ee
where $\nu_d$ is the correlation length exponent [$\nu_2=4/3; \
\nu_3=0.8\dots$]. From percolation theory (see e.g. Ref.~\onlinecite{Percolation}) it is also known that the probability of a given site to belong to a typical cluster scales as $\delta \lambda^{\beta_d}$ where $\beta_d$ is the infinite cluster density exponent [$\beta_2=5/36;\
\beta_3=0.39\dots$]. The clusters above and below the chemical potential are independent of each
other so that an estimate for the number of connected sites (not taking into account the memory effect) is
 \be
 N_c(\delta \lambda)= \frac{\left(\delta
     \lambda^{\beta_d}\right)^2}{\left(|\delta
     \lambda|^{\nu_d}\right)^d}.
 \label{Nc}
 \ee
According to our previous conclusions, the connection between sites (which give rise to the matrix elements $\Delta(\Upsilon_*,\Upsilon_{ij})$) requires that one of
the energies, either $\epsilon_i$ or $\epsilon_j$, is located within a band of the order of $T$ near the chemical potential.  This constraint implies that probability for connection between two sites, below and above the chemical potential, is of the order of
$T/E_{VRH}(T)=1/\Upsilon_c(T)$. Requiring that the full cluster formed by joining the clusters above and below the chemical potential (see Fig.~2) remains an infinite cluster leads to the condition $N_c(\delta \lambda)/\Upsilon_c(T)\approx 1$, which in turn implies
\be
\delta\lambda =\Upsilon_c(T)^{-1/\left({d\nu_d-2\beta}\right)}.
\ee
Substituting this result back into Eq.~(\ref{observablesigma2}), one obtains
 \be
\sigma \propto
\exp\left[-\Upsilon_c(T)+\alpha_d \Upsilon_c(T)^{\Delta_d}\right];\
\label{newsigma}
\ee
where
\be
\Delta_d=1-\frac{1}{d\nu_d-2\beta_d}, \label{Deltad}
\ee
and, in particular, $\Delta_2=\frac{25}{43}$ and $\Delta_3\simeq0.38$. The constant $\alpha_d$ is of order unity. It cannot be determined from the above considerations since the full statistical characteristics of the clusters near criticality are not universal (apart form the critical exponents), and depend on microscopic details of the system.
Equation \rref{newsigma} gives \req{Main-new} with
\be
\mu_d=\frac{\Delta_d}{d+1}.
\ee

\section{Conclusions.}
\label{sec5}

Formula  (\ref{Main-new}) describes, with exponential accuracy,  the temperature dependence of the variable-range hopping conductivity when memory effects are taken into account. Its derivation required two ingredients: The first one is the memory effect which leads to effective conductances of the resistors on the Miller-Abrahams network expressed by formula (\ref{ee}). This result implies that the conductivity in the variable-range hopping is described by the problem of two-color percolation. The second ingredient is the solution of the two-color percolation problem which is obtained by analyzing the critical behavior of the clusters size off-criticality. Equations (\ref{newsigma}) and (\ref{Deltad}) show how the critical exponents, which characterize the behavior near the percolation threshold, determine the temperature dependence of the subleading exponential term in (\ref{Main-new}).

Although this result has been obtained for spinless electrons, a similar calculation for spinful electrons yields the same result \cite{AA}. Moreover, since our arguments are general, we expect that a similar result holds also for systems with long range Coulomb interactions. The study of this case is left for future research.

An additional source of a different subleading term in the exponent of the hopping conductivity may come from the energy dependence of density of states. However this source is system dependent, and at sufficiently low temperatures should become negligible.

It is possible to misinterpret the subleading exponential contribution as an exponentially large prefactor. Assuming the preexponential factor in (\ref{Main-new}) to have a power-law from $B(T)= AT^m$, where $A$ is a $m$ are constants, and taking into account that $\mu_d$ is small,
\be
B(T) \exp\left[ \alpha_d\left(\frac{T_0}{T}\right)^{\mu_d} \right]  \approx A' T^{m-\alpha_d \mu_d}
\ee
where $A'=A T_0^{\alpha_d \mu_d} \exp(\alpha_d)$. This result might explain the difficulty of fitting the experimental data to formula (\ref{Main-Old}) with $B(T)= AT^m$ and an integer value of $m$. For instance, experimental results of the temperature dependence of the  variable-range conductivity in two dimensional samples\cite{Experiments1, Experiments2} were fit to formula (\ref{Main-Old}) with either $p=1/3$ or $p=1/2$ (depending whether Coulomb interactions can be neglected), and preexponential factor $B(T)= A T^{0.8}$.

\acknowledgements
We would like to thank M.~Gershenson, L.~Glazman,  A.~Klein, and Z.~Ovadyahu, for reading the manuscript prior to submission and for their comments. This research has been supported by the Simons foundation.
\appendix
\section{The short range memory approximation}

The purpose of this Appendix is to explain and justify the short range memory approximation used in order to derive Eq.~(\ref{ee}). In order to simplify the notations let us write Eq. (\ref{chargecorrelators-decoupled}) for the correlator, ${\cal C}_{ij}=\langle \langle \psi_i \psi_j\rangle\rangle$, in the concise form:
\be
\left[\eta_i\eta_j \frac{ d}{dt}- \hat{\cal D}_i-\hat{\cal D}_j+ \beta_{ij} \right]{\cal C}_{ij} = s_{ij}, \label{diffusion-like}
\ee
with
\be
\eta_j= n_j^{eq}(1- n_j^{eq}),
 \ee
 \be
 \beta_{ij}= \frac{(n_i^{eq}-n_j^{eq})^2}{\tau_{ij}},
 \ee
\be
s_{ij}=\frac{(n_i^{eq}-n_j^{eq})\langle \psi_j-\psi_i \rangle}{\tau_{ij}},
\ee
and
\be
\hat{\cal D}_i {\cal C}_{ij}= -\sum_{k\neq, ij}\frac{\eta_j}{\tau_{ik}}\left( {\cal C}_{kj} - {\cal C}_{i,j}\right).
\ee
The operator $\hat{\cal D}_i$ is a diffusion-like operator acting on the $i$th endpoint of the correlator.  Similarly, the operator $\hat{\cal D}_j$ acts on the $j$-th endpoint. Thus Eq.~(\ref{diffusion-like}) describes a diffusive-like motion of the two endpoints with a drain $\beta_{ij}$ and a source $s_{ij}$.

In what follow we consider typical transitions of variable-range hopping, i.e. transitions which involve sites whose occupation numbers at equilibrium are either very close to one ("full" sites) or very close to zero ("empty" sites). Furthermore, we focus our attention on the most relevant case in which the transition takes place from a full site to an empty site (or vice versa). The correlation, ${\cal C}_{ij}$, between sites of similar occupation numbers is negligible because the temporal change in these occupation numbers, due to hop of the hole or the electron, is exponentially small $\propto \exp(-E_{VRH}/T)$ [see also discussion below \req{ee}].

To begin with, let us first calculate the correction to the leading-order result [i.e. the solution of \req{charge-short-range}] within perturbation theory. To this end we rewrite \req{diffusion-like}, for the steady state solution, in a form that is suitable for solution by iterations:
\be
 \frac{{\cal C}_{ij}}{\chi_{ij}}= s_{ij} + \sum_{k\ne ij} \left[ \frac{\eta_j}{ \tau_{ik}} {\cal C}_{kj}+ \frac{\eta_i}{\tau_{jk}} {\cal C}_{ik} \right] \label{iterations}
\ee
where
\be
\frac{1}{\chi_{ij}}= \sum_{k\ne ij} \left[\frac{\eta_j}{\tau_{ik}} + \frac{\eta_i}{\tau_{kj}}\right] + \beta_{ij} \label{Mij}
\ee
Treating the sum on the right hand side of the equation as our perturbation, the zeroth order iteration gives:
\be
{\cal C}_{ij}^{(0)}= \chi_{ij}s_{ij}. \label{leading}
\ee
The next iteration shows that the leading correction to the above result comes from single hops of either the $i$-th endpoint or $j$-th endpoint of the correlator to some nearest neighbor site, $k$:
\be
{\cal C}_{ij}^{(1)}={\cal C}_{ij}^{(0)}
+ \chi_{ij}  \sum_{k\ne ij} \left[\frac{\eta_j}{\tau_{ik}} {\cal C}_{kj}^{(0)} + \frac{\eta_i}{\tau_{kj}} {\cal C}_{ik}^{(0)}\right]. \label{1st}
\ee
However, this correction is expressed as a sum over uncorrelated sources, $s_{lk}$, and therefore it represents a fluctuating quantity.

The next iteration gives:
\be
{\cal C}_{ij}^{(2)}\approx \left[1+ \chi_{ij} \sum_{k \ne ij} \left(\frac{\eta_j^2 \chi_{kj}}{\tau_{ik}^2} + \frac{\eta_i^2 \chi_{ki}}{\tau_{kj}^2}\right)  \right] {\cal C}_{ij}^{(0)}, \label{2nd}
\ee
where we have kept only the nonfluctuating contributions.

To find the magnitude of the correction to the leading result (\ref{leading}), we turn to estimate the factors $\chi_{ij}$. Taking into account that typical transitions on the Miller-Abrahams network are between sites which are either full or empty at equilibrium, we see that $\eta_k \sim e^{-E_{VRH}/T}$ is exponentially small for any $k$. Thus from \req{Mij} it follows that for a transition from a full site, $n_i^{eq} \approx 1$, to an empty site, $n_j^{eq} \simeq 0$, we have  $\chi_{ij} \approx 1/\beta_{ij}\sim \tau_{typical}$, where $\tau_{typical}$ is the typical hopping time. Now let us consider the additional factors which appear in \req{2nd}, $\chi_{kj}$ and $\chi_{ik}$. These factors are maximal for the case where the next hops are to sites of similar occupation numbers. In this case case $\chi_{ik}\sim \chi_{kj} \sim e^{(E_{VRH}/T)}\tau_{typical}$. From here we obtain that the correction to the zeroth-order result (\ref{leading}) is small by order of $e^{-E_{VRH}/T}$. The fluctuating contribution associated with the first order correction, (\ref{1st}), is small by the same order.

In order to complete the justification of the short range memory approximation, we turn to consider the contribution from long trajectories  associated with large number of hops of the two endpoints of the correlator, ${\cal C}_{ij}$. Our aim is to show that there is no divergent term coming from very long trajectories. We shall estimate this contribution for the most dangerous case in which hopping takes place between sites of similar occupation at equilibrium. We also take into account that, on the percolating network where transport takes place, closed loops are rare, and therefore the main contribution of long time dynamics comes from returning trajectories to both endpoints. The contribution from far away sources can be neglected because the probability that trajectories emanating from the $ij$ bond meet again at some remote bond, $kl$, is small (as closed loops are rare), and because these remote sources are independent and give a fluctuating contribution that in the long time limit self averages to zero. This implies that only the source and the drain associated with the $ij$ bond should be taken into account. It also means that the diffusive motion of the two endpoint is approximately independent, and therefore the total return probability after time $t$, is a product of the return probabilities of the two endpoints, where each one of them will be denoted by ${\cal P}(t)$. Hence, the contribution of the long time dynamics to the steady state solution of \req{diffusion-like} is:
 \be
{\cal C}_{ij}\approx \int_{0}^\infty dt e^{-\beta_{ij}t}{\cal P}^2(t) s_{ij}, \label{cijbar2}
\ee
where we have redefined the time as $\eta_i \eta_j t \to t$.

From percolation theory it is known that the return probability on the percolation cluster is
\be
{\cal P}(t) \sim \frac{1}{t^{d_s/2}}
\ee
where $d_s$ is the spectral dimension of the cluster. The latter is approximately $1.3$ both in two and in three dimensions\cite{Percolation2}, thus the integral (\ref{cijbar2}) rapidly converges in the limit $t \to \infty$, as $\beta_{ij} \sim 1/\tau_{typical}$, and is dominated by the short time limit.
 Notice that the integral converges in the upper limit, $t \to 0\infty$, even for $\beta_{ij}=0$, thus fluctuations in the transition time, $\tau_{ij}$, do not play an important role.

This result establishes the validity of the short range memory approximation, and confirms that the perturbative result (\ref{leading}) which is used in order to obtain \req{ee} holds with a correction of order $e^{- E_{VRH}/T}$.

\end{document}